\documentclass[showpacs,twocolumn,prl,floatfix,amssymb,aps]{revtex4}
\usepackage{graphicx}

\begin{document}

\hsize\textwidth\columnwidth\hsize\csname@twocolumnfalse\endcsname

\title{Low-energy excitations of the one-dimensional half-filled $SU(4)$ Hubbard model
with an attractive on-site interaction: Density-matrix renormalization-group calculations and
perturbation theory}
\author{Jize Zhao and Kazuo Ueda}
\affiliation{Institute for Solid State Physics, University of
Tokyo, Kashiwa, Chiba 277-8581, Japan}
\author{Xiaoqun Wang}
\affiliation{Department of Physics, Renmin University of China,
Beijing 100872, China}
\date{\today}
\begin{abstract}
We investigate low-energy excitations of the one-dimensional
half-filled $SU(4)$ Hubbard model with an attractive on-site
interaction $U<0$ using the density matrix renormalization group
method as well as a perturbation theory. We find that the ground
state is a charge density wave state with a long range order. The
ground state is completely incompressible since all the
excitations are gapful. The charge gap which is the same as the
four-particle excitation gap is a non-monotonic function of $U$,
while the spin gap and others increase with increasing $|U|$ and
have linear asymptotic behaviors.
\end{abstract}

\maketitle

The Hubbard model is one of the most classic models for strongly
correlated electronic systems and has attracted long-term interest
since the pioneering work in 1960s\cite{HUBB1}.  By taking the
on-site Coulomb interaction into account, it well explains the
puzzle that some materials with half-filled band are insulators.
However, this model which consider only a single band, on-site
Coulomb interaction and the nearest-neighbor hopping is often
thought being oversimplified. To account for other features beyond
the Mott physics\cite{MOTT}, there are two kinds of natural
extensions: one is to incorporate the orbital degree of freedom
which may be called multi-band Hubbard model\cite{IMADA1}, and the
other is to consider the hopping and/or the Coulomb interaction
with longer ranges.

In the last several years, ultra-cold atomic experiments evoke
systematic studies of correlation effects in the optical lattice
systems where interactions are tunable through Feshbach resonance.
The Hubbard model becomes again an appropriate one to envisage
some relevant issues with both positive and negative interactions.
Recently, fermionic atoms with higher spins are successfully
trapped into optical lattices \cite{MODUG1}. This calls for a
generalization of the $SU(2)$ Hubbard model into the $SU(N)$
case\cite{AFFLECK1,WU1}. In this paper, we study one-dimensional
$SU(4)$ Hubbard model which is represented as
\begin{equation}
\mathcal{H}=-t\sum_{i=1}^{L}\sum_{\sigma}(c^{\dagger}_{i\sigma}c_{i+1\sigma}+h.c.)+\\
\frac{U}{2}\sum_{i=1}^{L}\sum_{\sigma\ne\sigma^{'}}n_{i\sigma}n_{i\sigma^{'}}
\label{HAM}
\end{equation}
where $t>0$ is a hopping matrix element, $L$ the number of the
lattice sites, $\sigma$ and $\sigma^{'}$ the spin indices taking
$-\frac{3}{2}$, $-\frac{1}{2}$, $\frac{1}{2}$ and $\frac{3}{2}$.
$c^{\dagger}_{i\sigma}(c_{i\sigma})$ denotes
the creation(annihilation) operator of a particle with spin
$\sigma$ at the site $i$,
$n_{i\sigma}=c^{\dagger}_{i\sigma}c_{i\sigma}$ is the corresponding
number operator and $U$ is the on-site interaction.

This model is not exactly solvable even in one dimension in
contrast to the $SU(2)$ case\cite{LIEB1,CHOY1}. Nevertheless,
some aspects of physical properties can be reliably explored by
some analytical approaches as well as numerical methods such as
density matrix renormalization group (DMRG)\cite{WHITE1, PESCH1}
and Quantum Monte Carlo simulations\cite{ASSARAF1, ASSARAF2}. Most recently,
an $SO(8)$ symmetry regime was proposed between $0<U<3t$ at
half-filling by Assaraf et al\cite{ASSARAF1} with using a
nonperturbative renormalization group method and quantum Monte
Carlo simulation. They found that the low-energy spectrums are
gapful in this regime. The similar results were also shown later by
Szirmai and S\'olyom for the other $N>2$ case\cite{SZIRMAI1}.
Those studies were concentrated on the repulsive case $U>0$,
but for the $U<0$ case few results are obtained so far. On the
other hand, it is well known that the one-dimensional attractive
half-filled $SU(2)$ Hubbard model is described by a Luther-Emery
liquid\cite{LUTHER1}, in which the charge excitation is gapless,
whereas the spin excitation is gapful.  By the hidden $SU(2)$
transformation the $SU(2)$ Hubbard model with $U$ can be mapped to
the one with $-U$, while for the $SU(4)$ case such a mapping does not 
exist so that one cannot obtain any insights into the
low-energy properties through the mapping. In this paper, we will
show that the $SU(4)$ Hubbard model at half-filling with the
attractive interaction belongs to a different universality class
from the $SU(2)$ one.

Let us start with a perturbation theory for the strong coupling
regime. For this purpose, we rewrite the Hamiltonian (\ref{HAM})
as $\mathcal{H}=\mathcal{H}_{t}+\mathcal{H}_{u}$, where the
hopping term $\mathcal{H}_{t}=-t\sum_{i\sigma}
(c^{\dagger}_{i\sigma} c_{i+1\sigma}+h.c.)$ is regarded as a
perturbation and the on-site interaction
$\mathcal{H}_{u}=\frac{U}{2}\sum_{i\sigma\ne\sigma^{'}}
n_{i\sigma}n_{i\sigma^{'}}$ as the zeroth order Hamiltonian has
highly degenerate ground states in which each site is either fully
occupied by four particles forming a $SU(4)$ singlet, or empty. Up
to the second-order, the effective Hamiltonian is given by
\begin{equation}
\mathcal{H}^{(2)}_{eff}=\frac{2t^2}{3U}P\sum_{i}n_{i}P
-\frac{t^2}{6U}P\sum_{i}n_{i}n_{i+1}P,
\label{HEFF}
\end{equation}
where $P$ is a projection operator which projects a state onto the
subspace spanned by the ground states of $\mathcal{H}_{u}$, and
$n_{i}=\sum_{\sigma}n_{i\sigma}$ the number operator at the site
$i$. The hopping term $\mathcal{H}_{t}$ 
lifts the degeneracy of $\mathcal{H}_{u}$ and gives rise to both the 
energy gain of $\frac{4t^{2}}{3U}$ per site at half-filling as denoted by the
first term of (\ref{HEFF}) and an effective repulsive interaction between 
particles on the nearest-neighbor sites as denoted by the second 
term of (\ref{HEFF}). The second term induces essentially a
charge-density-wave (CDW) ground state with a true long range
order such that every other site is fully occupied with an empty
site in between, which is consistent with the mean field result for the 
weak coupling region\cite{WU1}. Moreover, ${\mathcal H}^{(2)}_{eff}$ has two-fold
degenerate ground states, each of which is a $SU(4)$ singlet.

An important question concerning the CDW ground state is whether
it is metallic or insulating. To address this question, we need to
examine charge excitations. The charge gap $\Delta_{c}$ is the
energy difference of the lowest excitation in the spin singlet
channel from the ground state as defined by
$\Delta_{c}=E_1(L,2L,0)-E_{0}(L,2L,0)$, in which $E_{n}(L,N,S)$
stands for the the $n-$th excitation energy in a spin-$S$ channel
with $L$ sites and $N$ particles. And another interesting issue is
to explore the relevance of the four-particle excitation gap
$\Delta_4$ to $\Delta_c$ for the $SU(4)$ symmetry. $\Delta_{4}$
represents the energy cost of adding four particles or holes into
the systems such that $\Delta_{4}=\frac 1
2\left[E_{0}(L,2L+4,0)\right.+E_{0}(L,2L-4,0)\left.-2E_0(L,2L,0)\right]$.
From ${\mathcal H}^{(2)}_{eff}$, one can easily find
$\Delta_{c}=-\frac{8t^{2}}{3U}$ based on the fact that the motion
of four particles from one of fully occupied sites to its neighbor
costs a minimal energy. Similarly, adding four particles to the
system would gain the energy $6U+2\times(-\frac{8t^2}{3U})$, while
adding four holes costs $-6U$, thus one has $\Delta_{4} =
-\frac{8t^{2}}{3U}$. It turns out that the four-particle
excitation gap is essentially the same as the charge gap and
finite.

However, the situation is completely different for the $SU(2)$
case, where the ground state is metallic. To understand this, one
can write the effective Hamiltonian in the strong coupling limit
for the $SU(2)$ case as\cite{EMERY1}:
\begin{eqnarray}
\mathcal{H}^{(2)}_{eff,su(2)} & = &
\frac{2t^2}{U}P\sum_{i\sigma}n_{i\sigma}P \label{HEFFSU2}\\
                              & -& \frac{t^2}{U}P\sum_{\langle{ij}\rangle\sigma}
             (n_{i\sigma}n_{j\sigma}-c^{+}_{i\sigma}c^{+}_{i\bar{\sigma}}
             c_{j\bar{\sigma}}c_{j\sigma})P\nonumber.
\end{eqnarray}
This Hamiltonian distinguishes itself from ${\mathcal
H}^{(2)}_{eff}$ with its last additional term which allows a
``pair hopping" process between the two neighbor sites. Although
the second repulsive term would be in favor of forming a CDW
ground state with gapful charge excitations, the extra ``pair
hopping" term eventually destabilizes this CDW long range order
resulting in gapless charge excitations. On the other hand for the $SU(4)$ 
case, similar hopping term can only appear after calculating to 
higher order. Therefore, the
perturbation theory in the strong coupling regime shed light on
the different nature of the ground states between the $SU(2)$ and
$SU(4)$ cases.

In order to go beyond the validity range of the above perturbation
theory for a full exploration of the low-energy properties, we
have performed systematic DMRG computations. There are actually
some difficulties inherent to this model, such as a large number
of degrees of freedom for each site and different edge states so
that some measures are taken necessarily to reach sufficient
accuracy in our computations. To calculate $\Delta_c$, we have to
use the periodic boundary condition (PBC) with the even number of
sites because of multi-edge excitations in the $SU(4)$ singlet
subspace when open boundary condition (OBC) is imposed. For other
gaps, we can efficiently expel the corresponding edge excitations
by an OBC algorithm. In particular, one lattice site is added at
each step and broken into two pseudo-sites. When the infinite
system algorithm is conducted with the size of the superblock up
to some odd number of sites $L$ which are preselected, the
sweeping procedure is performed for those $L$. The necessary
extrapolations for the thermodynamic limit are finally made
properly on the data with these preselected sites. In this case,
we have to redefine the gaps correspondingly, for instance
$\Delta_{4}$ $=E_0(L,2L+6,0) -E_0(L,2L+2,0)-6U$, where the
particle-hole symmetry is explicitly taken into account. In the
strong coupling region, PBC is often used to identify the bulk values of
a gap rather than edge-excitation energy obtained under OBC. In
our computations, $t$ is set to be unit and 2000 states are kept
for most cases and the maximal truncation error is the order of
$10^{-7}$.

\begin{figure}[ht]
\includegraphics[width=7.0cm,angle=0]{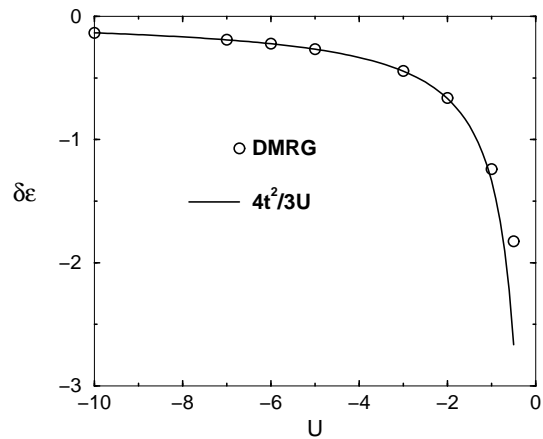}
\caption{Ground state energy correction per site from
$\mathcal{H}^{(2)}_{eff}$ (solid line) versus $U$ in comparison
with DMRG data ($\bigcirc$).} \label{fig1}
\end{figure}
Figure \ref{fig1} displays the ground state energy correction per site arising
from the hoping term for the thermodynamic limit. Both the
numerical results and the perturbation theory show this energy
 increases monotonically as $|U|$ increases and approaches to
 zero asymptotically. When $-2<U<0$, the results of the perturbation
theory deviates from DMRG ones, and the deviation becomes
significant for small $|U|$. This is reasonable because in this
region the hopping term $\mathcal{H}_{t}$ is no longer
perturbative. However, for $U<-2$, the perturbation theory
provides very good results which accurately agrees with the DMRG
data. On the other hand, our DMRG calculations with both OBC and
PBC show that for a finite and even $L$, the ground state is
unique and a $SU(4)$ singlet which belongs to the irreducible
representation $[1^4]$\cite{YAMASHITA1}. In addition, slightly
above the ground state there is one accompanied $SU(4)$ singlet
excited state, whose energy difference from the ground state
diminishes as $L\rightarrow\infty$. Therefore, one obtains
two-fold degenerate ground states in the thermodynamic limit,
which are consistent with our analysis based on ${\mathcal
H}_{eff}^{(2)}$. These degenerate CDW ground states with the long
range order result from the translational symmetry breaking.

\begin{figure}
\includegraphics[width=7.0cm,angle=0]{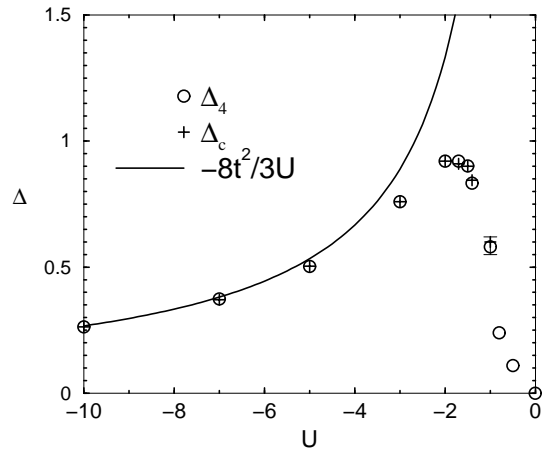}
\caption{DMRG results for charge gap (+) and four-particle gap
($\bigcirc$) versus $U$ in comparison with perturbation ones
(solid line). Error bars are smaller than size of symbols except for
$\Delta_c$ at $U=-1$ as estimated with keeping different
states.} \label{fig2}
\end{figure}

In FIG. \ref{fig2}, we show the DMRG results on the charge gap
$\Delta_c$ and the four-particle excitation gap $\Delta_4$ for the
entire range of $U<0$. First of all, one can see that both
$\Delta_c$ and $\Delta_4$ are non-vanishing for all finite $U<0$
so that the ground state is insulating rather than metallic in
contrast to the $SU(2)$ case. Secondly, these two gaps behave
non-monotonically with $U$. The maximum shown around $U=-2$
indicates a crossover region between weak and strong interaction
regimes. Third, the perturbation theory for the strong coupling
regime provides correctly the asymptotic behavior for large $|U|$
limit and shows a qualitative agreement with the DMRG results in
the strong coupling regime. The visible deviation from the DMRG
results sets on at about $U\approx-5$ lower than that ($U\approx
-2$) for the ground state energy correction shown in Fig. 1. In
the weak coupling regime, $\Delta_c$ and $\Delta_4$ shown by DMRG
decrease with increasing $U$. Finally, while ${\mathcal
H}^{(2)}_{eff}$ can predict $\Delta_c=\Delta_4$ only in the strong
coupling limit, our DMRG calculations show that within the
numerical accuracy $\Delta_c$ remains equal to $\Delta_4$ beyond
the strong coupling regime. Although it is difficult from the DMRG
calculations to obtain sufficiently accurate $\Delta_c$ for $-1
<U<0$ yet, it is reasonable to conclude that $\Delta_4$ is equal
to $\Delta_c$ for all $U<0$ in the $SU(4)$ case.

Now we turn to the other three types excitations: the first one is
the quasi-particle gap $\Delta_1$ for adding single particle or
single hole to the system, the second one spin gap $\Delta_s$
corresponding to the excitation energy in the spin triplet channel
from the ground state, and the last one two-particle gap
$\Delta_2$ defined as energy cost when two particles or two holes
are added to the system. While these three gaps together with
$\Delta_c$ and $\Delta_4$ essentially involve all kinds of
relevant excitations, they have significantly different behaviors.
Since the ground state is the CDW state with the long range order,
it is insightful to analyze those excitations in the Hartree-Fock
(HF) approximation. The CDW state in HF approximation can be
achieved by simply writing the on-site interaction as
$n_{i\sigma}n_{i\sigma^{'}}\approx \langle{n_{i\sigma}}\rangle
n_{i\sigma^{'}}+n_{i\sigma}\langle{
n_{i\sigma^{'}}}\rangle-\langle{n_{i\sigma}}\rangle\langle{
n_{i\sigma^{'}}}\rangle$ and assuming $\langle{n_{i\sigma}}\rangle
=n+(-1)^{i}\delta{n}$, where $n$ is the average number of
particles per site for each spin and $\delta{n}$ the corresponding
order parameter. At half-filling, one has $n=\frac{1}{2}$ and
$0\le{\delta{n}} \le\frac{1}{2}$. By further introducing
$a_{l\sigma}=c_{2l\sigma}$ and $b_{l\sigma}=c_{2l+1\sigma}$ for
each sublattice of the bipartite lattice, respectively, and taking
the Fourier transformation, then we can write down the
Hartree-Fock Hamiltonian as:
\begin{eqnarray}
\mathcal{H}^{HF} & = & -t\sum_{k\sigma}\left((1+e^{-ik}) a^{+}_{k\sigma}
b_{k\sigma}+h.c.\right) \nonumber \\
                 &   & +3U\delta{n}\sum_{k\sigma}(a^{+}_{k\sigma}a_{k\sigma}
                 -b^{+}_{k\sigma}b_{k\sigma})+const.
\label{HF-K}
\end{eqnarray}
Diagonalizing this Hamiltonian, one can obtain two bands for each
spin species $\sigma$ with the quasi-particle dispersions
$w^{\pm}_{\sigma}=\pm{ \sqrt{\Delta_1^2+4t^{2}\cos^{2}
{\frac{k}{2}}}}$ where $\Delta_1=-3U\delta{n}$. Moreover, one has
$\Delta_c=\Delta_s=2\Delta_1$ in the HF approximation. These gaps
can be then evaluated after solving the following  self-consistent
equation $2\delta{n}=\langle{a^{+}_{l\sigma}a_{l\sigma}} \rangle
-\langle{b^{+}_{l\sigma} b_{l\sigma}}\rangle$ for the order
parameter $\delta n$. In order to calculate the two-particle
excitation gap, however, it is necessary to account for the
particle-particle correlations into
the HF approximation, which is nothing but the random phase
approximation (RPA). For this purpose, one first constructs the
basis with two-particle excitations from the ground state of
${\mathcal H}^{HF}$ as follows:
\begin{eqnarray}
|\Psi_{pp'\sigma}\rangle=\alpha^{+}_{p\sigma}\alpha^{+}_{p'\bar{\sigma}}
|\Psi_{g}\rangle,   ~~~~~~
|\Psi_g\rangle=\prod_{k\sigma}\beta^{+}_{k\sigma}|0\rangle
\label{basis})
\end{eqnarray}
where $\alpha^{+}_{p\sigma}$ is an operator creating one
quasi-particle with momentum $p$ and spin $\sigma$ in the $w^+$
band and in $|\Psi_{g}\rangle$ the band $w^-_{\sigma}$ is fully
filled up by quasi-particles $\beta^{+}_{k\sigma}$ with the
momenta $k$ and spin $\sigma$. Then $\Delta_{2}$ can be obtained
by diagonalizing ${\mathcal H}$ on the above basis (\ref{basis}).
\begin{figure}[h!]
\includegraphics[width=7.4cm,angle=0]{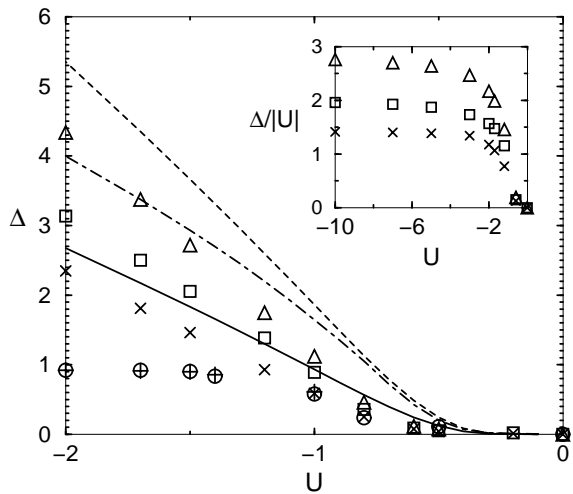}
\caption{DMRG results for one particle gap $\Delta_{1}$
($\times$), two-particle gap $\Delta_{2}$($\Box$), spin gap
$\Delta_{s}$ ($\triangle$), four-particle excitation gap
$\Delta_{4}$ ($\small\bigcirc$) and charge gap $\Delta_{c}$($+$)
are shown as a function of $U$. The results of HF-RPA for
$\Delta_1$, $\Delta_2$ and $\Delta_s$ denoted by solid,
 dot-dashed, and dashed lines, respectively. Inset shows
 $\Delta/|U|$ versus $U$ for $\Delta_1$ ($\times$), $\Delta_2$ ($\Box$) and
 $\Delta_s$ ($\triangle$).}
  \label{fig3}
\end{figure}

As compared to the DMRG results, we found that HF-RPA can provide
a qualitatively correct description for $\Delta_{1}$, $\Delta_2$,
and $\Delta_{s}$. Figure \ref{fig3} shows the DMRG data on all
five gaps as well as $\Delta_1,\Delta_2$, and $\Delta_s$ from
HF-RPA approximation for the region of $-2\leq U\leq 0$ and the
inset illustrates $\Delta_{1}$, $\Delta_2$, and $\Delta_{s}$ by
showing ratios for them over $|U|$ up to $|U|=10$. In contrast to
$\Delta_c$ and $\Delta_4$ as seen from Fig. 2, $\Delta_{1}$,
$\Delta_2$, and $\Delta_{s}$ increase with increasing $|U|$ and
become linear in large $|U|$ limit. It turns out that the relation
$\Delta_c=\Delta_s$, given by the HF approximation, is invalid for
general $U<0$. Moreover, it is unclear but beyond the present
approaches whether there is a symmetry enlargement similar to the
one proposed for the repulsive case\cite{ASSARAF1}. Nonetheless,
HF-RPA presents precise asymptotic behaviors for $\Delta_1$,
$\Delta_2$ and $\Delta_s$. In the weak coupling limit, the
exponential opening of these gaps can be well reproduced from the
solution to the self-consistent equation $\delta n\sim{-\frac
{2\pi t}{3U}e^{\frac{2\pi t}{3U}}}$ and with taking into account the
two-particle correlations. In the strong coupling limit, one has
$\Delta_1\sim -1.5U$ and $\Delta_s\sim-3U$ from $\delta
n\rightarrow0.5$, and $\Delta_2\sim-2U$ from the HF-RPA
calculations\cite{note}. The corresponding coefficients are in
good agreement with the DMRG results as can be seen from the
inset. On the other hand, the results of HF-RPA deviate from the
DMRG data apparently in the intermediate coupling regime, but this
is quite understandable since correlations involved in (\ref{HAM})
cannot be accurately handled in HF-RPA when $\mathcal{H}_{t}$ and
$\mathcal{H}_{u}$ become comparable, i.e. neither of them are
perturbative.

In summary, we have studied the low energy properties of the
one-dimensional half-filled $SU(4)$ Hubbard model with the
attractive on-site interaction by using the DMRG method as well as
the perturbation theory. We found that the ground state is a CDW
insulating state with the long range order in which the
translational symmetry is broken and all kinds of excitations are
gapful for finite $U<0$. Within our numerical accuracy, we found
that the four-particle excitation gap is the same as the charge
gap. While the charge gap (the four particle excitation gap)
behaves non-monotonically, the others increase with increasing
$|U|$ and have a linear-U asymptotic behavior with different
coefficients. Therefore, we believe that the one-dimensional
attractive half-filled Hubbard model for the $SU(4)$ and $SU(2)$
cases belong to different universality classes. Moreover, we find
that the nature for the $SU(4)$ case can be further generalized to
the other $SU(N>2)$ cases\cite{ZHAO1}. At the end, it is
worthwhile to mention that since the four-particle excitation gap
as well as the charge gap are the smallest energy scale for the
$SU(4)$ case with $U<0$, it would be very interesting to detect
four-particle process (excitations) in an ultra-cold fermionic
atom system with the hyperfine spin-3/2.

We would like to thank Y. Yamashita and Y.Z. Zhang for fruitful
discussions. J. Zhao acknowledges J. S\'olyom for the helpful
correspondence. X.Wang is supported under the Grants 2005CB32170X
and NSFC10425417.

\vfill

\begin{references}
\bibitem{note} The asymptotic behavior of  $\Delta_1$, $\Delta_2$ and $\Delta_s$ can be obtained
               alternatively from $H_u$. For instance, $\Delta_1$ is given by considering that
               adding one hole costs energy $-3U$ and adding one particle gains no energy.
\end{references}
\end{document}